\documentclass[floats,aps,preprint]{revtex4}
\usepackage{mathrsfs}
\usepackage{graphicx}

\DeclareGraphicsExtensions{.jpeg,.pdf,.eps,.gif,.tif,.sxd}
\DeclareGraphicsExtensions{.pdf}

\begin{document}

\title{Probing anomalous relaxation by coherent multidimensional optical
spectroscopy }
\author{ Franti\v{s}ek \v{S}anda $^{*}$ and  Shaul Mukamel $^{\dagger}$ }
\address{$^{*}$
Charles University, Faculty of Mathematics and Physics,  Ke Karlovu 5,
Prague, 121 16 Czech Republic \\
$^{\dagger}$ Department of Chemistry, University of California, Irvine, CA 92697-2025 
}
\date{\today}

\begin{abstract}
\widetext\bigskip

We propose to study the origin of algebraic decay of two-point correlation
functions observed in glasses, proteins, and quantum dots by their nonlinear
response to sequences of ultrafast laser pulses. Power-law spectral
singularities and temporal relaxation in two-dimensional correlation
spectroscopy (2DCS) signals are predicted for a continuous time random walk
model of stochastic spectral jumps in a two level system with a power-law
distribution of waiting times $\psi (t)\sim t^{-\alpha -1}$. Spectroscopic
signatures of stationary ensembles for $1<\alpha <2$ and aging effects in
nonstationary ensembles with $0<\alpha <1$ are identified. \vfill E-mail: 
$^{\ast }$sanda@karlov.mff.cuni.cz, $^{\dagger }$ smukamel@uci.edu
\end{abstract}

\maketitle


Exponential relaxation profiles and correlation functions observed in
systems coupled to a bath with a short correlation time are signatures of
fast memory loss. Such normal relaxation can be described by Markovian (e.g.
Langevin, Fokker-Planck and master) equations of motion; Multipoint Green's
functions may then be factorized into products of two point functions and
carry no additional information. However, correlation functions in many
systems characterized by a complex free-energy landscape with a broad
distribution of dynamical barriers \cite{frauenfelder}, may acquire long
stretched-exponential or algebraic tails \cite{metzler}. Anomalous
variations of fluorescence emission (blinking) has been observed in
biomolecules \cite{xie}, polymers \cite{amblard}, quantum dots 
\cite{verberk,nesbitt,bawendi} and glasses. The fluorescence trajectories are
commonly analyzed using a continuous time random walk (CTRW) model of
spectral diffusion. 
In this letter we demonstrate how spectral lineshapes obtained from
two-dimensional correlation spectroscopy (2DCS) \cite{mukamel} may be used
to probe complex anomalous relaxation processes in the condensed phase. 2DCS
techniques are femtosecond optical analogues of NMR, that have been
successfully employed towards the study of the structure of peptides \cite{hochstrasser}, 
chemical exchange in liquids \cite{fayer} and exciton
migration in photosynthetic antennae \cite{fleming}. 
2DCS provides a bulk alternative to single molecule measurements of multipoint correlation
functions \cite{xie} useful for testing microscopic dynamic model. Connection to
anomalous relaxation requires a consistent theory of 2DCS signals with
Nonmarkovian spectral fluctuations, which is the goal of this letter.

The CTRW model is defined by a waiting time probability density function
(WTDF) $\psi (t)$ for stochastic jumps between various states. All memory is
erased at the time of the jump. This renewal (resetting) property makes it
possible to compute all statistical measures even in the absence of a
Markovian description for the probability distribution, and provides a
convenient formalism for describing long-term memory effects. When $\psi (t)$
has an exponential form $\psi (t)=\exp{(-t/\kappa _{1})}$ 
(where $\kappa_{1}\equiv \int_{0}^{\infty }t\psi (t)dt$ is mean waiting time) the system
may be described by a Makovian master equation, and the relaxation is
normal. However, models with a long time algebraic decay $\psi (t)\sim
t^{-\alpha -1}$, show anomalous phenomena at long times when the second
moment of $\psi (t)$ diverges $0<\alpha <2$. We consider a random walk which
is observed starting at time 0. The WTDF of the first jump $\psi ^{\prime
}(t)$ may differ from $\psi (t)$ since it depends on how the system was
prepared before $t=0.$ It must be treated with care, since it strongly
affects the nature of the anomalous ensemble. Stationary processes must
satisfy microscopic reversibility which implies that $\psi ^{\prime }(t)$
is given by a product of the survival probability $\phi (t)=\int_{t}^{\infty
}\psi (t^{\prime })dt^{\prime }$ that no jump had occurred for time $t$ and
the equilibrium density of jumps $1/\kappa _{1}$ resulting in $\psi ^{\prime
}(t)=\phi (t)/\kappa _{1}$ \cite{sanda1}.  A stationary CTRW is thus only
possible for $1<\alpha <2$ where the first moment $\kappa _{1}$ is finite.
The resulting power-law relaxation is slow, but eventually the system
reaches an equilibrium \cite{shlesinger}. For $0<\alpha <1$, $\kappa _{1}$
diverges. The system never equilibrates and shows aging effects (i.e.
dependence on the initial observation time). Aging phenomena are commonly
studied by preparing a nonstationary ensemble where all particles are
assumed to make a jump at the time origin, so that $\psi ^{\prime }(t)=\psi
(t)$. Signatures of aging were observed in fluorescence blinking of single
CdSe quantum dots, analyzed within CTRW and yielded $\alpha \approx 0.5$ 
\cite{brookmann,nesbitt,bawendi}. In contrast, the anomalous multipoint
correlations observed in fluorescence trace of conformation dynamics of
flavin proteins \cite{xie} showed symmetries due to microscopic
reversibility indicative of a stationary process.

We consider a system undergoing anomalous stochastic jumps between two
states $a$ and $b$. The system is further coupled to a two-level chromophore
with a ground $|g\rangle $ and an excited state $|e\rangle $ and dipole
moment $\mu _{eg}$, causing its transition frequency $\Omega _{eg}$ to
undergo stochastic fluctuations $\delta \Omega _{eg}(t)$. $\delta 
\Omega_{eg}$ can assume two values $\Omega _{0}$ and $-\Omega _{0}$ corresponding
to the system in state $a$ and $b$ respectively. This is known as the two
state jump (TSJ) model of spectral lineshapes \cite{barkai}. Observable
quantities are obtained by averaging over the ensemble of stochastic paths
of $\delta \Omega _{eg}(t)$.

We propose to probe this complex dynamics through the four-wave-mixing
signal generated by the coherent response of the chromophore to three short
optical pulses with wavevectors $\mathbf{k_{1}}$, $\mathbf{k_{2}}$ and 
$\mathbf{k_{3}}$ at times 0,$t_{1}$,$t_{1}+t_{2}$. The signal
detected at time $t_{1}+t_{2}+t_{3}$ is described by the nonlinear response
function which depends on the three experimentally controlled parameters 
$t_{1},t_{2},t_{3}$. The photon echo signal generated in the 
$\mathbf{k_{I}=-k_{1}+k_{2}+k_{3}}$ phase-matching direction is given by \cite{principles} 
\begin{eqnarray}
\mathscr{S}_{I}(t_{3},t_{2},t_{1}) &=&2(i/\hbar )^{3}\mu _{eg}^{4}\theta
(t_{3})\theta (t_{2})\theta (t_{1})e^{i\Omega _{eg}(\eta
t_{1}-t_{3})} \nonumber \\
&\times&e^{-\Gamma (t_{1}+t_{3})} {\bigg\langle}\exp \left[ {\
-i\int_{t_{1}+t_{2}}^{t_{1}+t_{2}+t_{3}}\delta \Omega _{eg}(\tau )d\tau }
\right]
 \exp \left[ {\ i\eta \int_{0}^{t_{1}}\delta \Omega _{eg}(\tau )d\tau 
}\right] {\bigg \rangle}  \label{k_1}
\end{eqnarray}
with $\eta =1$. We have added a homogenous dephasing rate $\Gamma $. We
shall display this signal by 2D $\omega _{1},\omega _{3}$ correlation plots 
\begin{equation}
S_{I}(\omega _{1},t_{2},\omega _{3})=-Im\int \int 
\mathscr{S}_{I}(t_{1},t_{2},t_{3})e^{i(\omega _{1}t_{1}+\omega _{3}t_{3})}dt_{1}dt_{3}
\label{imshape}
\end{equation}
%
where $t_{2}$ is held fixed. A different signal, $\mathscr{S}_{II}$, in the 
$\mathbf{k_{II}=k_{1}-k_{2}+k_{3}}$ direction is similarly given by setting 
$\eta =-1$. We further considered the combination 
$S_{A}(\omega _{3},t_{2},\omega _{1})\equiv S_{I}(\omega _{3},t_{2},-\omega
_{1})+S_{II}(\omega _{3},t_{2},\omega _{1})$ 
which shows absorptive peaks \cite{tokmakoff}. An optical coherence is
encoded in the system during $t_{1}$ and probed during $t_{3}$. By varying
the delay $t_{2}$ between these periods (which is typically much longer than 
$t_{1}$ and $t_{3}$) we can explore correlations between dynamical events of
the stochastic system. Multipoint correlation functions ordinarily obtained
in single molecule spectroscopy can thus be obtained from bulk measurements.

Microscopic reversibility in stationary ensembles implies that 
$\mathscr{S}_{I}(t_{3},t_{2},t_{1})=\mathscr{S}_{I}^{\ast }(t_{1},t_{2},t_{3})$ and $\mathscr{S}_{II}(t_{3},t_{2},t_{1})=\mathscr{S}_{II}(t_{3},t_{2},t_{1})$,
resulting in the following symmetry of the lineshape 
\begin{equation}
S_{A}(\omega _{3},t_{2},\omega _{1})=S_{A}(\omega _{1},t_{2},\omega _{3}).
\label{sym_diagonal}
\end{equation}
When the system has no memory during the $t_{3}$ interval regarding its
state during $t_{1}$ $S_{I},$ and $S_{II}$ may be factorized as 
$S_{I}(t_{3},t_{2},t_{1})=2(i/\hbar )K(t_{3})K^{\ast }(t_{1})$ , and 
$S_{II}(t_{3},t_{2},t_{1})=2(i/\hbar )K(t_{3})K(t_{1})$. \ Here $K(t)\equiv
(i/\hbar )\mu _{eg}^{2}e^{(-\Gamma +\Omega eg)t}\langle \exp
[-i\int_{0}^{t}\delta \Omega _{eg}(\tau )d\tau ]\rangle $, is the linear
response function. The correlation signal reduces in this case to the
product of the linear absorption $W_{A}(\omega )\equiv $ $Im\int_{0}^{\infty }K(t)\exp [i\omega t]dt$ lineshapes 
\begin{equation}
\hbar S_{A}(\omega _{3},t_{2}\rightarrow \infty ,\omega _{1})=4W_{A}(\omega
_{1})W_{A}(\omega _{3}).  \label{factorization}
\end{equation}
Algebraic memory will result in a slow convergence to this asymptotic
lineshapes. In addition, the spectra will diverge at certain frequencies
where the factorization (Eq. (\ref{factorization})) does not hold, as will
be shown below.

The response functions will be calculated by introducing a $2\times 2$
matrix $\hat{G}_{\nu }$ in $a,b$ space whose $jl$ element gives the
contribution of paths with initial state $l$ and final state $j$ by
averaging over $l$ and summing over $j$. 
\begin{equation}
\mathscr{S}_{\nu }(t_{3},t_{2},t_{1})=\sum_{jl}\left( G_{\nu }\right)
_{jl}(t_{3},t_{2},t_{1})\left[ \rho \right] _{l}(t=0)  \label{defmult}
\end{equation}%

In the case of Markovian relaxation, $G_{\nu }$ is given by a product of
three Green's functions representing the time evolution during the $t_{1}$, 
$t_{2}$ and $t_{3}$ intervals. These Green's functions can be calculated by
solving the stochastic Liouville equation, which combines the Liouville
equation for coherence evolution with a rate equation for the jumps 
\cite{tanimura}. The anomalous four-point $\hat{G}_{\nu }$, in contrast, may not
be factorized in this manner. Calculating it requires some bookkeeping of
jump events. For each of the three time intervals $t_{1},t_{2},t_{3}$ we
distinguish between two possibilities; either there was no jump or there was
at least one jump during that interval. According to this classification, 
$\hat{G}_{\nu }$ is given by a sum of 8 terms $\hat{G}_{\nu }^{m}$,
($m=1,\ldots ,8$). $\hat{G}_{\nu }^{m}$ are constructed by convoluting matrix
factors for propagation between the first and the last jump in the $t_{i}$
interval (if any jump occurred in $t_{i}$) and factors for the coherent
evolution between the consecutive jumps in different intervals (last in
earlier and first in later) \cite{sanda1}. The calculation is conveniently
made in Laplace domain where convolutions become simple multiplications and
the integral equation for the propagator factor becomes an algebraic
equation. Making use of the causality of the response functions (Eq.(\ref{k_1})), 
the 2D lineshapes (Eq.\ref{imshape}) were obtained by analytical
continuation of the Laplace domain response functions calculated in Appendix
D of Ref. \cite{sanda1} by setting $s_{1}=\Gamma +i\eta (\omega _{1}-
\Omega_{eg})$, and $s_{3}=\Gamma -i(\omega _{3}-\Omega _{eg})$, where $s_{j}$ is
the Laplace variable conjugated to $t_{j}$.

We first consider a stationary ensemble with anomalous relaxation where \cite{sanda1}: 
\begin{equation}
\tilde{\psi}_{W}(s)=\frac{1}{1+\kappa _{1}s/\left[ 1+(\kappa _{\alpha
}s)^{\alpha -1}\right] };\quad 1<\alpha <2;  \label{weak_anomalous}
\end{equation}
$\tilde{\psi}(s)$ is the Laplace transform of $\psi (t)$. This WTDF has a
mean waiting time $\kappa _{1}$ and the long time 
$t>>(\kappa _{1}\kappa_{\alpha }^{\alpha -1})^{1/\alpha }$ 
asymptotics $\psi _{W}(t)\sim \kappa_{\alpha }^{\alpha -1}\kappa _{1}/t^{\alpha +1}$ . 
In all plots we use the dimensionless frequency units 
$(\omega _{j}-\Omega _{eg})/\Omega _{0}$ by setting 
$\Omega _{eg}=0,\Omega _{0}=1.$

No jumps occur during $t_{1}$ and $t_{3}$ in the slow fluctuation $\Omega
_{0}\kappa _{1}>>1$ limit. The linear lineshape has two peaks at frequencies 
$\omega =\pm 1$, and the 2D spectrum for short delay $t_{2}\sim 0$ consists
of two diagonal peaks (the system is in the same state during $t_{1}$ and 
$t_{3}$) at $(\omega _{1},\omega _{3})$ (1,1) and (-1, -1) centered at these
two frequencies. For normal relaxation the peaks are Lorentzian. The
anomalous model shows the same peak pattern, but the peaks are divergent due
to long-tailed correlations. For $\Gamma =0$ the linear absorption peaks
diverge as $\Delta \omega ^{\alpha -2}$ where $\Delta \omega _{1}\equiv
\omega _{1}-\Omega _{eg}-\Omega _{0}$ \cite{sanda1,barkai}.

In Fig 1 we display the $S_{I}$, $S_{II}$ and $S_{A}$ signals for slow
fluctuations at $t_{2}=0$, and $\Gamma =0$. All panels show prominent two
diagonal peaks at $(1,1)$ and (-1,-1). $S_{A}$ is simpler due to
interference of $S_{I}$ and $S_{II}$, which have opposite signs in specific 
$\Delta \omega _{1},\Delta \omega _{3}$ quadrants, and substantially cancel.
Along the $\omega _{1}=\pm 1$ and $\omega _{3}=\pm 1$ lines, $S_{I}$ and 
$S_{II}$diverge as $\sim \Delta \omega _{3}^{\alpha -2}$ and $\sim \Delta
\omega _{1}^{\alpha -2}$ respectively. $S_{A}$ is finite along 
$\omega_{1}=\pm 1$ (except for the $\omega _{3}=\pm 1$ peaks), but is
nondifferentiable with respect to $\omega _{1}$. The $\omega _{3}=\pm 1$
line has the same analytical structure.

The most rapidly divergent term $\hat{G}_{\nu }^{8}$ which represents the
paths that has no jump during the entire time $t_{1}+t_{2}+t_{3}$ is given
by (for the $(1,1)$ diagonal peak) 
\begin{eqnarray}
S_{A}^{8}(\omega _{1},t_{2}=0,\omega _{3}) &=&\frac{4\mu _{eg}^{4}}{\hbar
^{3}}{\bigg[}\frac{\Delta \omega _{3}}{\Delta \omega _{3}^{2}-\Delta \omega
_{1}^{2}}Im\tilde{\phi}^{\prime }(\Gamma -i\Delta \omega _{3})  \nonumber \\
&+&\frac{\Delta \omega _{1}}{\Delta \omega _{1}^{2}-\Delta \omega _{3}^{2}}
Im\tilde{\phi}^{\prime }(\Gamma -i\Delta \omega _{1}){\bigg]}  \label{pd}
\end{eqnarray}
For $\Gamma >0$ Eq.(\ref{pd}) is regular, for $\Gamma =0$ it diverges as 
$S_{A}\approx \frac{4\mu _{eg}^{4}}{\hbar ^{3}}\kappa _{\alpha }^{\alpha
-1}\sin {\left[ \pi (2-\alpha )/2\right] }\Delta \omega _{3}^{\alpha -3}$ 
($\Delta \omega _{3}\rightarrow 0$,$\Delta \omega _{1}=0$). This is
illustrated in the linear log-log plot in Fig 2A showing in $S_{A}$ vs.
$\omega_{3}$ for $\omega _{1}=1$. The $\alpha -3$ exponent is shown in the inset.

For fast fluctuations $\Omega _{0}\kappa _{1}<<1$ (not shown), the state of
the system rapidly changes during the $t_{1}$,$t_{3}$ intervals. For normal
relaxation, both 1D and 2D lineshapes then consist of a single Lorentzian
peak at the average frequency (motional narrowing). For anomalous dynamics
the survival probability in the initial state is substantial even for fast
fluctuations and the two peak divergencies of the slow limit (Fig 1) are
still retained. An additional central peak at $(0,0)$ shows up.

The variation of $S_{A}$ with delay time $t_{2}$ in the slow fluctuation
limit is displayed in Fig 3. We see a buildup of (finite, as shown at right
panel) off-diagonal cross-peaks, whose lineshapes are dominated by paths
where the system is in a different state during $t_{1}$ and $t_{3}$. The
(-1,1) peak represents paths where the system was in state $a$ during $t_{1}$
and $b$ during $t_{3}$. The $(1,-1)$ peak represents the reverse sequence.
Note that this 2DCS equilibrium measurement records subensembles of
trajectories without having to perturb the system or perform a single
molecule measurement. Contour elongation of both diagonal and cross peaks
along the $\omega _{1,3}=\pm 1$ directions is a signature of the long-time
memory and $S_{A}$ may not be factorized as in Eq. (\ref{factorization}) at
these frequencies. 
Outside these regions, or for finite $\Gamma $ the asymptotic lineshape Eq. 
(\ref{factorization}) is approached algebraically as $t_{2}$ is increased. By
including a finite dephasing rate $\Gamma ,$ peak divergencies are cured and
the lineshapes do not differ significantly from the normal relaxation case.
However, the population evolution during $t_{2}$ is not affected by
dephasing and the cross-peak dynamics is still anomalous. To study how the 
$(1,-1)$ cross peak grows to its long time value (Eq. (\ref{factorization}))
we have considered the quantity $Z(t_{2})\equiv |S_{A}(\omega
_{1},t_{2},\omega _{3})-S_{A}(\omega _{1},\infty ,\omega
_{3})|/|S_{A}(\omega _{1},0,\omega _{3})-S_{A}(\omega _{1},\infty ,\omega
_{3})|$ for $\omega _{1}=1$, $\omega _{3}=-1$. Straight lines in log-log
plots of $Z(t_{2})$ vs. $t_{2}$ (Fig 2B) indicate algebraic relaxation. The
asymptotic exponent (slope in Fig 2B) approaches $\alpha -1$, which is
identical to that of the frequency correlation function $\langle \delta
\Omega _{eg}(t)\delta \Omega _{eg}(0)\rangle $ \cite{sanda1}.

Finally we discuss aging effects for $0<\alpha <1$ case, where $\kappa _{1}$
diverges. This describes nonergodic nonstationary processes which never
equilibrate, and their properties change with time \cite{klafter}. We model
it by a random walk which is started by a jump at some fixed time. The
response function depends on the time elapsed from the start of a random
walk $t_{in}$ to the first laser pulse at 0 \cite{barbi,aquino}. This effect
is fully described by allowing $\psi ^{\prime }(t;t_{in})$ to depend on 
$t_{in}$. In Laplace space we find $\tilde{\psi}^{\prime }(s;s_{in})=
[\tilde{\psi}(s_{in})-\tilde{\psi}(s)]\tilde{\psi}(s_{in})/[1-\tilde{\psi}
(s_{in})](s-s_{in})$ \cite{long}. The nonlinear lineshapes measured in the
time-domain provide a direct measure of response function, unlike
frequency-domain measurements which involve stationarity in order to connect
with response function \cite{barkai}. However, due to the lack of
equilibration, averaging over consecutive pulse sequences may depend on the
repetition rate: A fresh ensemble with the same $\psi ^{\prime }(t)$ must be
prepared before each application of the pulse sequence for the response
function to be physically meaningful.

We have calculated the response functions for the model 
$\tilde{\psi}_{N}(s)=1/[1+(\kappa s)^{\alpha }]$, where $\psi (t)\sim (\kappa
/t)^{1+\alpha }$ and $0<\alpha <1$.  In Fig 1B the $S_{A}$ lineshape is
shown for $t_{in}=0$, i.e. the random walk is started each time the first
pulse interacts with the sample and $\psi ^{\prime }(t)=\psi (t)$ \cite{klafter}. 
Microscopic reversibility breaks down for nonstationary processes
and obviously the symmetry (Eq. (\ref{sym_diagonal})) is violated. The
nonstationary anomalous ensembles starts with some jump rate which depends
on the initial preparation. At long times it approaches the equilibrium
value $1/\kappa _{1}$, which vanishes for the present model \cite{barbi}.
The higher rate during the (earlier) $t_{1}$ interval compared to $t_{3}$
results in broader peaks along $\omega _{1}$ axis (compared to $\omega _{3}$). 
Contour elongation along the $\omega _{1,3}=\pm 1$ axes is again observed
with divergent diagonal peaks of complex structure. Our simulations
demonstrate that two-dimensional correlation plots of the signals obtained
from the response of the system to sequences of multiple laser pulses carry
specific and direct signatures of complex dynamics.

The support of the M\v{S}MT \v{C}R (MSM 0021620835), NSF (CHE-0446555) and NIH (2RO1GM59230-05) is
gratefully acknowledged.

\newpage

\textbf{\Large Figure captions}\newline

\begin{description}
\item {Fig 1} (Color Online) (A): The 2D $S_{I}(-\omega _{1},\omega _{3})$
(left), $S_{II}(\omega _{1},\omega _{3})$ (middle) and $S_{A}(\omega
_{1},\omega _{3})$ (right) signal (Eq.(\ref{k_1})) for the WTDF 
(Eq.(\ref{weak_anomalous})) at $t_{2}=0$,for $\alpha =1.2$, and $\kappa _{\alpha
}/\kappa _{1}=0.25$, $\Omega _{0}\kappa _{1}=2.0$, $\Omega _{0}=1$,
$\Omega_{eg}=0$. (B): The 2D $S_{A}$ signal for the nonstationary random walk 
$\alpha =0.5$, $t_{2}=0$, $t_{in}=0$ $\kappa \Omega _{0}=2.0$, $\Omega _{0}=1$,
 $\Omega _{eg}=0$.

\item {Fig 2} (Color online) Panel (A): Peak divergence along $\Delta \omega
_{1}=0$. $t_{2}=0$, $\alpha =1.2$(dotted), $1.5$ (dashed), and $1.8$
(solid). Other parameters are the same as in Fig 1. Inset: peak exponent 
$\gamma \equiv \frac{d\log {\left[ S_{A}(\Delta \omega _{1}=0,t_{2}=0,\Delta
\omega _{3})\right] }}{d\log {[\Delta \omega _{3}]}}$. Panel (B): $Z(t_{2})$
for the WTDF (Eq. (\ref{weak_anomalous})) $\kappa _{\alpha }/\kappa _{1}=0.5$
, $\Omega _{0}\kappa _{1}=1.0$, $\alpha =1.2$(dotted), $\alpha =1.5$
(dashed), and $\alpha =1.8$ (solid), $\Omega _{0}=1$. Power-law growth of
cross peaks is seen.

\item {Fig 3} (Color Online) Variation of the 2D $S_{A}$ signal (Eq.(\ref{k_1}))
for the WTDF (Eq. (\ref{weak_anomalous})) with the time delays (left
to right) $t_{2}=0$, $2\kappa _{1}$, $10\kappa _{1}$. Other parameters same
as Fig (1A). Right panel: Cross peak for $t_{2}=10\kappa _{1}$ on an
expanded scale.
\end{description}

\newpage
\begin{center}
\scalebox{0.90}[0.90]{\includegraphics{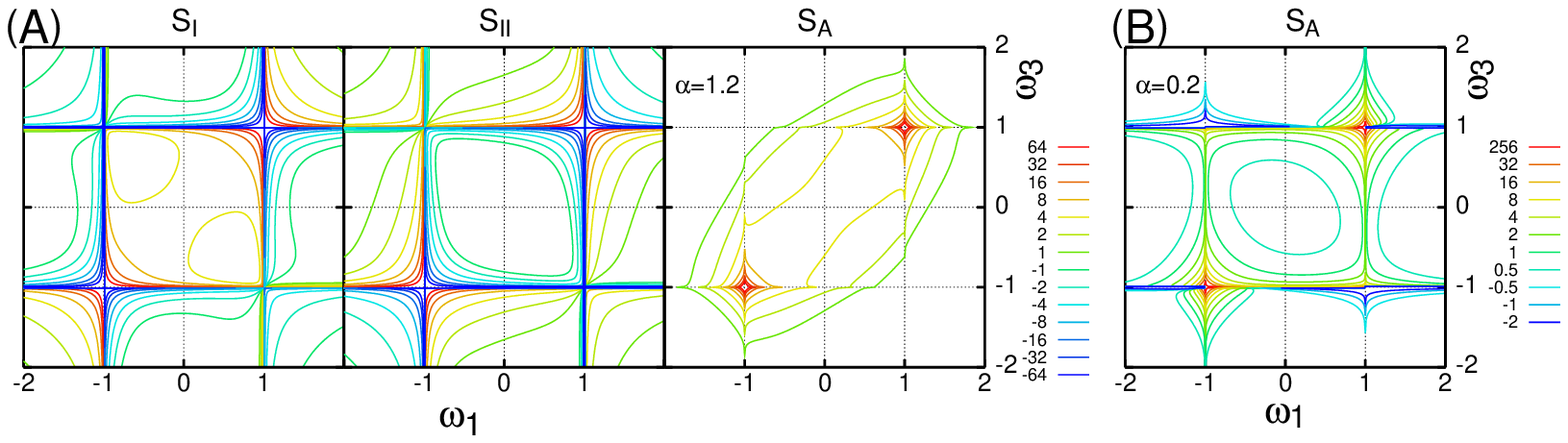}}
\end{center}
$\quad \quad \quad  \quad \quad\quad \quad \quad \quad \quad$   {\large \bf Fig 1}

\newpage

\begin{center}
\scalebox{1.0}[1.0]{\rotatebox{0}{\includegraphics{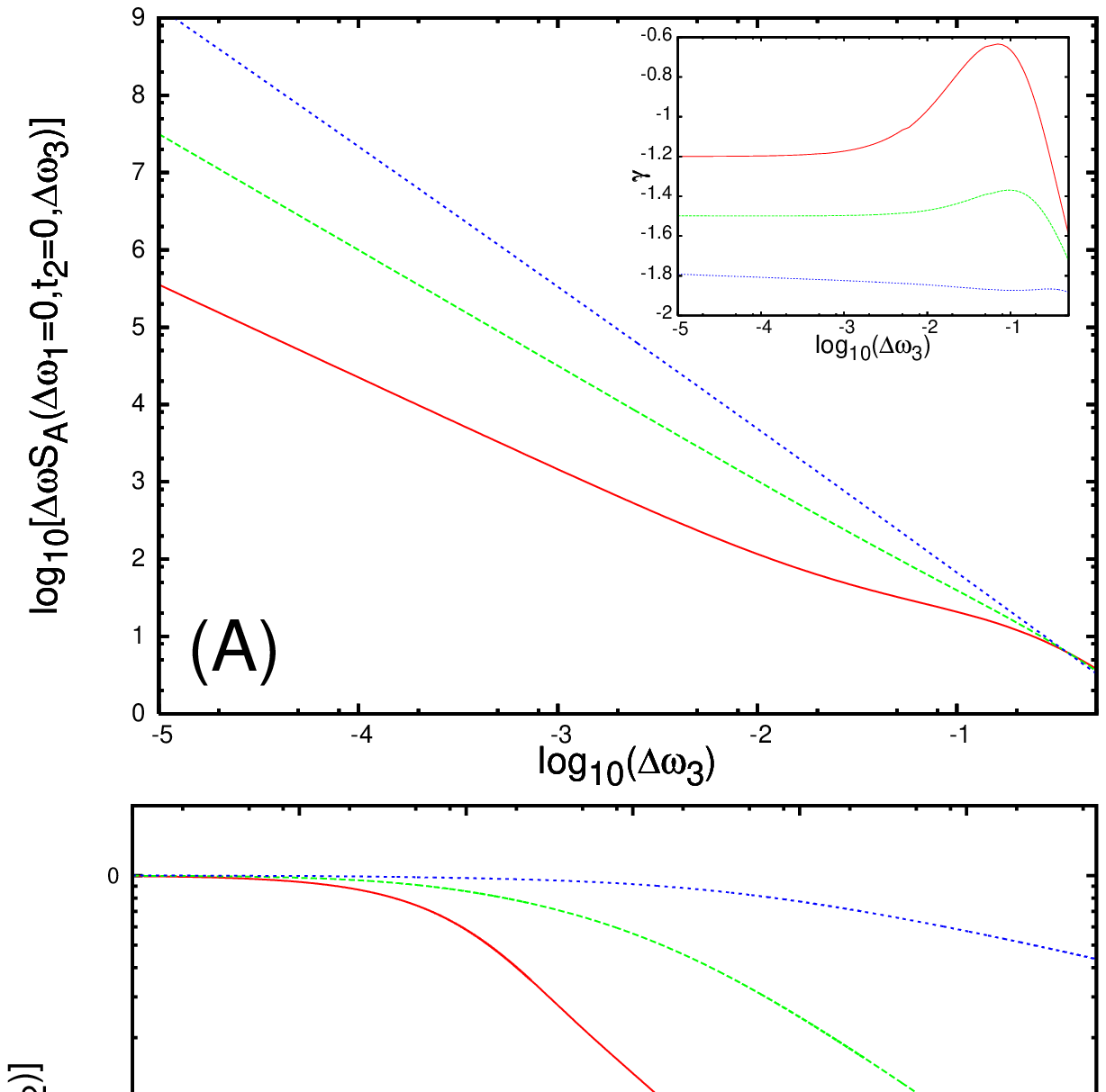}}}
\end{center}
\vspace{220pt}
{\large \bf Fig 2}

\newpage
\begin{center}
\scalebox{0.80}[0.80]{\rotatebox{0}{\includegraphics{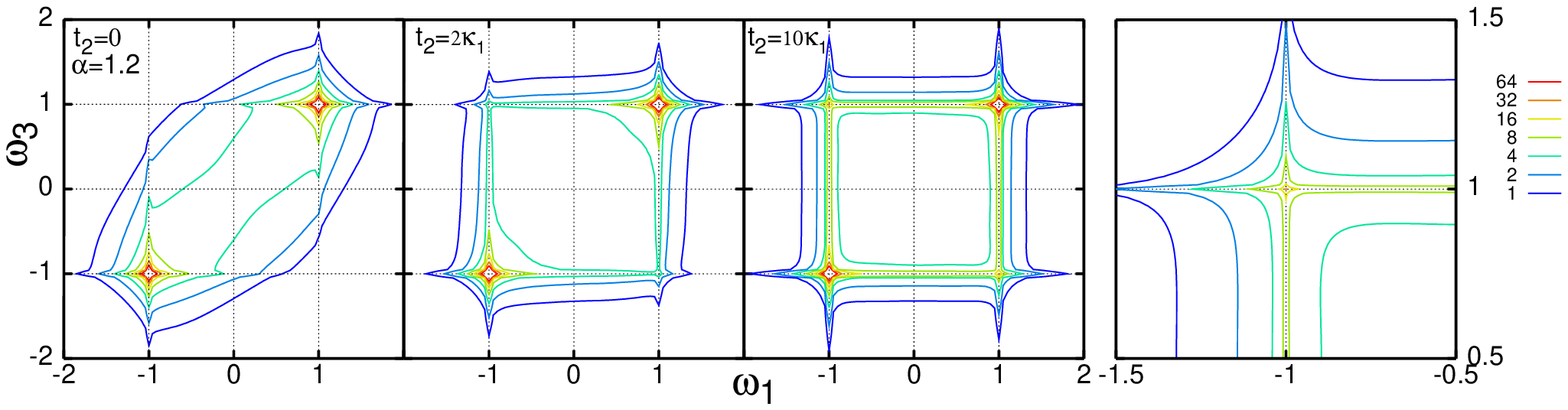}}}
\end{center}

{\large \bf Fig 3}
\end{document}